\pdfoutput=1

\documentclass[aps,prl,twocolumn,superscriptaddress,10pt]{revtex4-1}

\usepackage{amsmath,amsfonts,amssymb}
\usepackage{graphicx}

\newcommand{\bra}[1]{\ensuremath{\left\langle#1\right|}}
\newcommand{\ket}[1]{\ensuremath{\left|#1\right\rangle}}

\DeclareMathOperator{\Tr}{Tr}

\begin{document}

\title{Coherent control via weak measurements in $^{31}$P single-atom electron and nuclear spin qubits}
\author{J. T. Muhonen}
\altaffiliation[Current address: ]{Center for Nanophotonics, AMOLF, Science Park 104, 1098 XG Amsterdam, The Netherlands}
\email{muhonen@amolf.nl}
\affiliation{Centre for Quantum Computation and Communication Technology, School of Electrical Engineering and Telecommunications, UNSW Australia, Sydney NSW 2052, Australia}

\author{J. P. Dehollain}
\altaffiliation[Current address: ]{QuTech \& Kavli Institute of Nanoscience, TU Delft, 2628 CJ Delft, The Netherlands}
\affiliation{Centre for Quantum Computation and Communication Technology, School of Electrical Engineering and Telecommunications, UNSW Australia, Sydney NSW 2052, Australia}

\author{A. Laucht}
\affiliation{Centre for Quantum Computation and Communication Technology, School of Electrical Engineering and Telecommunications, UNSW Australia, Sydney NSW 2052, Australia}

\author{S. Simmons}
\altaffiliation[Current address: ]{Department of Physics, Simon Fraser University, Burnaby BC V5A 1S6, Canada}
\affiliation{Centre for Quantum Computation and Communication Technology, School of Electrical Engineering and Telecommunications, UNSW Australia, Sydney NSW 2052, Australia}

\author{R. Kalra}
\altaffiliation[Current address: ]{School of Mathematics \& Physics, University of Queensland, Brisbane QLD 4072, Australia.}
\affiliation{Centre for Quantum Computation and Communication Technology, School of Electrical Engineering and Telecommunications, UNSW Australia, Sydney NSW 2052, Australia}

\author{F. E. Hudson}
\affiliation{Centre for Quantum Computation and Communication Technology, School of Electrical Engineering and Telecommunications, UNSW Australia, Sydney NSW 2052, Australia}

\author{D. N. Jamieson}
\affiliation{Centre for Quantum Computation and Communication Technology, School of
Physics, University of Melbourne, Melbourne VIC 3010, Australia}

\author{J. C. McCallum}
\affiliation{Centre for Quantum Computation and Communication Technology, School of
Physics, University of Melbourne, Melbourne VIC 3010, Australia}

\author{K. M. Itoh}
\affiliation{School of Fundamental Science and Technology, Keio University, 3-14-1 Hiyoshi,
223-8522, Japan}

\author{A. S. Dzurak}
\affiliation{Centre for Quantum Computation and Communication Technology, School of Electrical Engineering and Telecommunications, UNSW Australia, Sydney NSW 2052, Australia}

\author{A. Morello}
\email{a.morello@unsw.edu.au}
\affiliation{Centre for Quantum Computation and Communication Technology, School of Electrical Engineering and Telecommunications, UNSW Australia, Sydney NSW 2052, Australia}

\begin{abstract}
The understanding of weak measurements and interaction-free measurements has greatly expanded the conceptual and experimental toolbox to explore the quantum world. Here we demonstrate single-shot variable-strength weak measurements of the electron and the nuclear spin states of a single $^{31}$P donor in silicon. We first show how the partial collapse of the nuclear spin due to measurement can be used to coherently rotate the spin to a desired pure state. We explicitly demonstrate that phase coherence is preserved throughout multiple sequential single-shot weak measurements, and that the partial state collapse can be reversed. Second, we use the relation between measurement strength and perturbation of the nuclear state as a physical meter to extract the tunneling rates between the $^{31}$P donor and a nearby electron reservoir from data, conditioned on observing no tunneling events. Our experiments open avenues to measurement-based state preparation, steering and feedback protocols for spin systems in the solid state, and highlight the fundamental connection between information gain and state modification in quantum mechanics.
\end{abstract}

\date{\today}
\maketitle

The quantum measurement postulate, as found in quantum mechanics textbooks, implicitly describes projective (von Neumann) measurements, where a measurement apparatus is coupled to a quantum system and, upon performing the measurement, returns a unique value $a_k$ for some observable $\hat{A}$ of the quantum system. If the system was initially in the state $\ket{\psi}$, the act of measurement leaves it in the state $\ket{\phi_k}$, the eigenstate of the observable $\hat{A}$ with eigenvalue $a_k$. The non-deterministic and non-unitary process through which the act of measurement transforms the initial state $\ket{\psi}$ into the final state $\ket{\phi_n}$ is known as ''wavefunction collapse'', and has been the subject of a century of debate and controversy.

However, as was already appreciated by von Neumann \cite{Neumann1932}, the projective measurement is only a limiting case. One can also have a detector which is only partially correlated with some observable of the quantum system, and therefore returns only partial information on the system state. Accordingly, the wavefunction needs not be fully projected onto an eigenstate, but is only weakly disturbed by the measurement process. The implications and applications of such ``weak measurements'' and corresponding partial collapse of the quantum state have gained considerable attention, especially in the context of quantum information processing. Recent experiments on superconducting qubits have demonstrated partial wavefunction collapse \cite{Katz2006}, measurement reversal \cite{Katz2008}, stabilized Rabi oscillations using quantum feedback \cite{Vijay2012}, direct observation of quantum trajectories \cite{Murch2013,Weber2014}, reduction of decoherence via ``uncollapsing'' \cite{Zhong2014}, and observation of the back-action steering from a variable strength measurement \cite{Hatridge2013,Groen2013}.

In this Letter, we describe how to apply the principles of weak quantum measurements to the electron and nuclear spin states of an individual $^{31}$P donor atom in silicon. In the context of quantum measurement, the $^{31}$P atom provides access to many key features, in particular related to negative-result measurements \cite{Dicke1981} and quantum steering \cite{Wiseman2007,Cavalcanti2016}. In particular, we show that weak measurements can be used to \emph{phase coherently} control the state of the $^{31}$P nuclear spin, and that it is possible to preserve phase coherence through the sequential measurement and control steps. This aspect of weak measurements has not been explicitly clarified in the recent literature, and opens avenues to measurement based state preparation, EPR steering and possible feedback protocols in these systems. As a further demonstration of the applicability of the weak measurement toolbox to the $^{31}$P system, we show how the tunneling rate of the electron to a nearby electron reservoir can be extracted from a dataset conditioned on having no tunneling events, in a spirit similar to the Elitzur-Vaidman bomb \cite{Elitzur1993}.

\begin{figure}
\includegraphics[width=0.42\textwidth]{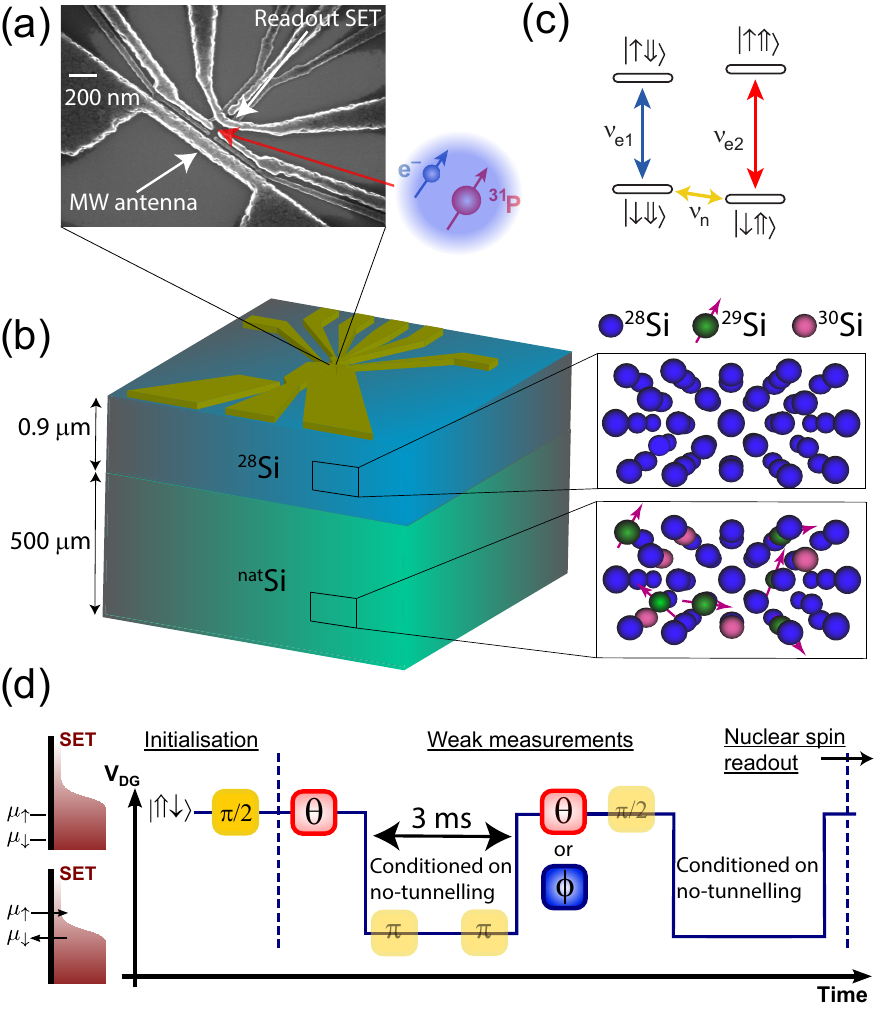}
\caption{
(a) Scanning electron micrograph of a device identical to the one used in the experiment.  A broadband microwave antenna is used to provide both nuclear and electron spin resonance pulses, and a single-electron transistor (SET) detects electron tunneling events in real time. 
(b) Schematic of the device layout, highlighting the isotopically enriched $^{28}$Si epilayer on top of a natural silicon substrate. 
(c) Energy diagram of the electron-nuclear spin system, with labels for the transition frequencies relevant to the present experiments. 
(d) Schematic of the pulse sequence used in the measurements. The solid line represents the combined effect of the voltage of the electrostatic gates ($V_{\rm DG}$) adjusting the chemical potential of the donor electron with respect to the SET island, schematically shown on the left. Blue or red boxes represent ESR pulses at $\nu_{\rm e1}$ or $\nu_{\rm e2}$, yellow boxes NMR pulses at $\nu_{\rm n}$. The semi-transparent boxes are only needed for tomography ($\sigma_x$ and $\sigma_y$ components). They are, from left to right: two refocusing pulses (around y-axis) and one phase-modulated pulse to define the tomography axis. The timing of the pulse lengths is not to scale.
} \label{device}
\end{figure}

Figure~\ref{device}(a) shows a scanning electron microscope image of our device, which is fabricated on an isotopically enriched $^{28}$Si substrate \cite{Itoh2014}, and where the $^{31}$P atom is introduced via single-ion implantation \cite{Donkelaar2015}. This system has gained considerable attention in the field of solid-state quantum information processing, since it contains two natural qubits (the electron spin, with $S=1/2$ and basis states $\ket{\uparrow}, \ket{\downarrow}$, and the $^{31}$P nucleus, with spin $I=1/2$ and basis states $\ket{\Uparrow}, \ket{\Downarrow}$) that exhibit extremely long coherence times \cite{Tyryshkin2012,Saeedi2013,Muhonen2014}, high quantum gate fidelities \cite{Muhonen2015,Dehollain2016b} and can be efficiently entangled with each other \cite{Simmons2011,Dehollain2016a}. 

At its core, the quantum state of the $^{31}$P system is accessible through the measurement of the $z$-projection of the electron spin, where $z$ is the axis along which a strong external magnetic field $B_0$ ($\approx 1.5$~T in the present experiment) is applied. The donor is placed in close proximity ($\approx 25$~nm \cite{Mohiyaddin2013}) to a cold ($T \approx 100$~mK) electron reservoir. Under suitable biasing conditions, the donor-bound electron can tunnel into the cold reservoir if and only if it is in the excited $\ket{\uparrow}$ state. The positively charged donor left behind after this tunneling event shifts the bias point of a nearby single-electron transistor (SET) and switches it to a high conductance state. Conversely, a $\ket{\downarrow}$ electron cannot escape the donor, leaving the SET in a near-zero conductance state. This spin-dependent tunneling process \cite{Elzerman2004,Morello2009,Morello2010} thus gives rise to a single-shot measurement, with fidelity in excess of 90\% \cite{Morello2010}. This mechanism provides a near-ideal negative-result measurement for the $\ket{\downarrow}$ state, which is identified by the absence of a signal in the SET current. 

The $^{31}$P nuclear spin couples to the electron through the Fermi contact hyperfine interaction $A \mathbf{I} \cdot \mathbf{S}$, with $A \approx 97$~MHz in this specific device. As a consequence, the electron spin can have two possible resonance frequencies, $\nu_{\rm e1,2} = \gamma_{\rm e}B_0 \mp A/2$ [Fig.~\ref{device}(c)], where $\gamma_{\rm e} \approx 28$~GHz/T is the electron gyromagnetic ratio. Single-shot nuclear readout \cite{Pla2013} is obtained by initializing the $\ket{\downarrow}$ state and applying a microwave $\pi$-pulse at e.g. $\nu_{\rm e1}$, where subsequently measuring the electron $\ket{\uparrow}$ state indicates that the nuclear spin state was $\ket{\Downarrow}$. Since we work in the limit $\gamma_{\rm e}B_0 \gg A$, the hyperfine interaction can be approximated with $A I_z S_z$, and therefore commutes with the $S_z$ electron spin observable. This means that the readout of the $z$-projection of the nuclear spin is of quantum nondemolition type \cite{Braginsky1980}, and can be repeated to achieve a readout fidelity approaching 99.9\% \cite{Pla2013}, well beyond that of a single-shot electron readout. The use of an electron $\pi$-pulse is just a limiting case, where one gains maximum information about the nuclear spin state. Here instead we explore the more general case where the electron rotation angle is $\theta \neq \pi$ \cite{Blok2014}, which causes the subsequent electron readout to provide only partial information on the nuclear state. This realizes a tunable weak measurement, with strength controlled by the electron rotation angle $\theta$. We show below that, as a result of a weak nuclear measurement conditioned on measuring electron \ket{\downarrow}, the nuclear state can be coherently rotated to an arbitrary pure state. This could be extended to provide an interesting implementation of EPR steering \cite{Wiseman2007,Cavalcanti2016} with spins in the solid state, by applying ESR pulses simultaneously on both $\nu_{\rm e1}$ and $\nu_{\rm e2}$ \cite{supp}. Our experiments were conducted by exciting only one ESR frequency at a time, and therefore we will refrain from using the term ``steering'' to describe the process.  

Let us assume that the nuclear spin is initially in the state $\ket{\psi_{\rm n0}} = (\ket{\Downarrow} + \ket{\Uparrow})/\sqrt{2}$, while the electron spin is initialized in its ground state $\ket{\downarrow}$. We then apply a microwave pulse at frequency $\nu_{\rm e2}$ to produce a rotation by an angle $\theta$ of the electron spin, conditioned on the nuclear spin being in the $\ket{\Uparrow}$ state. The full electron-nuclear state then becomes $\ket{\Psi_{\rm en}} = \left[ \ket{\Downarrow \downarrow} + \cos(\theta/2) \ket{\Uparrow\downarrow} + \sin(\theta/2) \ket{\Uparrow \uparrow} \right]/\sqrt{2}$. A readout of the electron spin state will then produce $\ket{\uparrow}$ with probability $P_{\uparrow}= \sin^2(\theta/2)/2$ and leave the nuclear spin state $\ket{\Uparrow}$. More interestingly, with probability $P_\downarrow = [1+\cos^2(\theta/2)]/2$ the electron readout will produce $\ket{\downarrow}$ and leave the nuclear spin in a coherent superposition state $\ket{\psi_{\rm n}} = [\ket{\Uparrow}+\cos(\theta/2)\ket{\Downarrow}]/[1+\cos^2(\theta/2)]$, which has therefore been rotated from the original state $\ket{\psi_{\rm n0}}$ using only electron spin resonance (ESR) pulses and electron spin measurements. Importantly, as we show below, this rotation is fully coherent and can be used the prepare any nuclear spin superposition state. The rotation is probabilistic in the sense that it can fail (if the outcome of the electron readout is $\ket{\uparrow}$), but in case of a success (heralded by the $\ket{\downarrow}$ electron readout) the steered state is fully deterministic.

A more complete description of the process is obtained through a density matrix formalism \cite{supp}. The initial nuclear spin state is
\begin{equation}
\rho_0 = \ket{\psi_{\rm n0}}\bra{\psi_{\rm n0}} = \frac{1}{2}
\begin{bmatrix}
    1 & 1  \\
    1 & 1
\end{bmatrix}.
\label{eq:superposition}
\end{equation}
After the $\theta$ rotation of the electron spin (initially $\ket{\downarrow}$) conditioned on the $\ket{\Uparrow}$ nuclear state, and a $\ket{\downarrow}$ electron readout, the nuclear spin is left in the state
\begin{equation}
\rho(\theta) = \frac{1}{1+\cos^2(\theta/2)}
\begin{bmatrix}
    \cos^2(\theta/2) & \cos (\theta/2)  \\
    \cos (\theta/2) & 1
\end{bmatrix}, \label{rho}
\end{equation}
which notably is a pure state for all values of $\theta$. This readily generalizes to multiple electron rotation and measurement steps. For example, after two sequential applications of the sequence, the nuclear spin state is (conditional on reading $\ket{\downarrow}$ at both steps)
\begin{eqnarray}
\rho(\theta_1,\theta_2) &=& \frac{1}{1+\cos^2(\theta_1/2)\cos^2(\theta_2/2)} \\ 
&\times &
\begin{bmatrix}
    \cos^2(\theta_1/2)\cos^2(\theta_2/2) & \cos (\theta_1/2)\cos (\theta_2/2) \\
    \cos (\theta_1/2) \cos (\theta_2/2) & 1
\end{bmatrix},
\nonumber
\end{eqnarray}
assuming phase coherence is preserved at the intermediate electron readout step (see below).

An interesting scenario appears if the second electron rotation is applied at $\nu_{\rm e1}$ instead of $\nu_{\rm e2}$, so that the rotation is conditioned on the nuclear $\ket{\Downarrow}$ state. Calling $\phi$ the rotation angle of the microwave pulse at $\nu_{\rm e1}$, the final state becomes
\begin{eqnarray}
\rho(\theta,\phi) &=& \frac{1}{\cos^2(\phi/2)+\cos^2(\theta/2)} \\ 
&\times &
\begin{bmatrix}
    \cos^2(\theta/2) & \cos (\theta/2)\cos (\phi/2) \\
    \cos (\theta/2) \cos (\phi/2) & \cos^2(\phi/2)
\end{bmatrix}.
\nonumber
\end{eqnarray}
If we set $\phi = \theta$, the final state is $\rho(\theta,\theta) = \rho_0$. This is known as ``measurement reversal'' \cite{Korotkov2006,Katz2008}: the second weak measurement of the nuclear spin erases the effect of the first one. 

\begin{figure} [tb]
\includegraphics[width=0.46\textwidth]{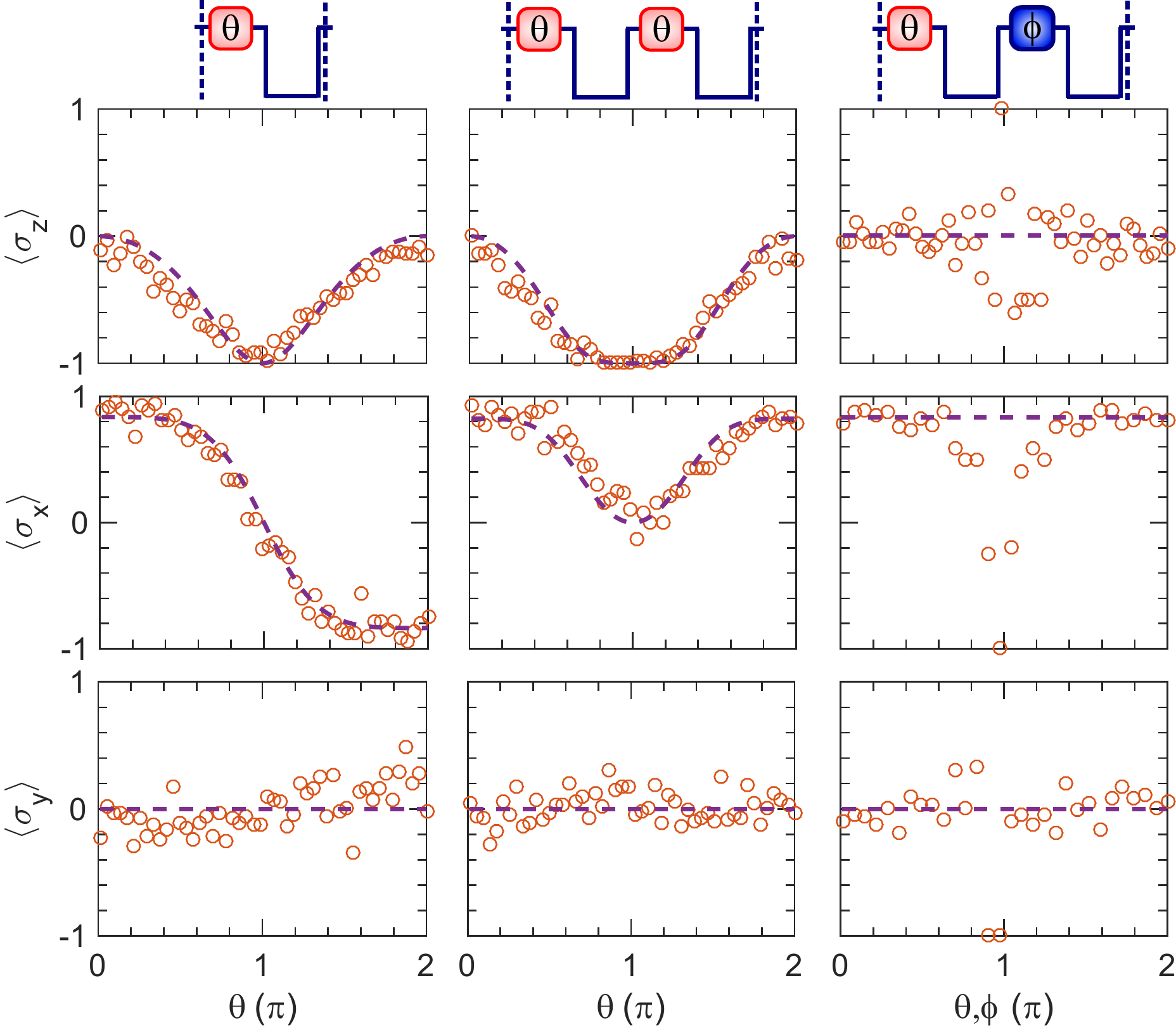}
\caption{Quantum control of a nuclear spin with electron spin resonance pulses, observed through quantum state tomography of $\sigma_z$ (1$^{\rm st}$ row), $\sigma_x$ (2$^{\rm nd}$ row) and $\sigma_y$ (3$^{\rm rd}$ row) as a function of the electron spin rotation angle $\theta$ on the ESR frequency $\nu_{\rm e2}$. On the columns, from left to right: one weak measurement, two weak measurements (each with rotation angle $\theta$), and measurement reversal (rotation by $\theta$ on $\nu_{\rm e2}$ and then by $\phi$ on $\nu_{\rm e1}$, here $\phi=\theta$). Circles: experimental data. Lines: calculated results from the density matrices as explained in the main text, without any free fitting parameters. The lines for $\sigma_x$ have been scaled by a constant to match the asymptotic values, which are not exactly unity due to rotation errors in the tomography pulse.} \label{steering}
\end{figure}

Figure~\ref{steering} shows experimental data obtained with full quantum state tomography, i.e. measurement of all three nuclear spin components $\sigma_z = (\rho_{1,1}-\rho_{2,2})$, $\sigma_x = (\rho_{1,2}+\rho_{2,1})$, and $\sigma_y = (\rho_{1,2}-\rho_{2,1})$. 

The left column of Fig.~\ref{steering} is the result of a single nuclear rotation step, consisting of an ESR pulse at $\nu_{\rm e2}$ inducing a rotation of angle $\theta$ around the $x$-axis to the electron spin state, followed by single-shot electron readout, and postselection on the $\ket{\downarrow}$ outcome. The dashed lines, in excellent agreement with the data, show the expected nuclear state, on the basis of the density matrix description presented above, without any free fitting parameters.

The middle column in Fig.~\ref{steering} illustrates the application of two sequential rotation steps, conducted for simplicity with the same ESR rotation angle $\theta$ on $\nu_{\rm e2}$ at both steps. The fact that the data (especially the $\sigma_x$-component) follows the theoretical predictions indicates that the nuclear state remains coherent throughout the sequence, which contains two weak nuclear measurements. In other words, the partial collapse of the nuclear state after the first weak measurement is a phase coherent, predictable process, although the evolution is non-unitary. A minimum requirement for observing this effect is that the dephasing time of the nuclear spin qubit has to be longer than the electron readout time. The $^{31}$P nuclear spin qubit in $^{28}$Si already has an intrinsically long dephasing time ($T_{2}^{\ast} \approx 0.5$~ms \cite{Muhonen2014}), but here we further extend it by applying two NMR refocusing pulses during the 3 ms electron readout step (see Fig. 1(d)). We also frequency-modulate the NMR source to track the resonance frequency of the nuclear spin qubit during the electron readout phase, since the change in the donor electrostatic potential under readout conditions causes a Stark shift of the resonance frequency \cite{Laucht2015}. 

On the right column of Fig.~\ref{steering} we present the so-called measurement reversal \cite{Korotkov2006,Katz2008}, which requires a rotation by $\theta$ on $\nu_{\rm e2}$ and rotation by $\phi = \theta$ on $\nu_{\rm e1}$. As predicted, we recover the original state each time (again, conditional on obtaining $\ket{\downarrow}$ at each electron readout step). Note that when $\theta = \pi$, the nuclear measurement becomes fully projective and the probability of a successful reversal becomes zero (all success probabilities are presented in supplementary Fig. 1). The data points around $\theta = \pi$ are thus only statistical fluctuations. 

We now explore the possibility of performing a weak electron spin measurement, and the effects that such a measurement has on the nuclear spin. The spin-dependent tunneling mechanism that provides a discrimination between the $\ket{\uparrow}$ and $\ket{\downarrow}$ states yields a fully projective measurement only in the limit $\Gamma_{\uparrow,\mathrm{out}} t_{\rm m} \rightarrow \infty$, where $t_{\rm m}$ is the measurement time and $\Gamma_{\uparrow,\mathrm{out}}$ is the tunnel-out rate for a $\ket{\uparrow}$ electron, defined such that the probability for a $\ket{\uparrow}$ electron to have tunnelled out of the donor after time $t_{\rm m}$ is $P_{\uparrow,\mathrm{out}}(t_{\rm m}) = 1 - \exp(-\Gamma_{\uparrow,\mathrm{out}} t_{\rm m})$. For a finite value of $\Gamma_{\uparrow,\mathrm{out}} t_{\rm m}$, the absence of a tunnel-out event constitutes only a weak $\ket{\downarrow}$ measurement. 

The effect on the nuclear spin of a weak electron measurement can be captured quantitatively in the density matrix formalism, by modifying Eq.~\ref{rho} to include the probability $1-P_{\uparrow,\mathrm{out}}(t_{\rm m})$ that an $\ket{\uparrow}$ does not tunnel out within the measurement time \cite{supp}.

\small
\begin{eqnarray}
\rho(\theta,t_{\rm m}) &=& \frac{1}{1+\cos^2(\theta/2)+(1-P_{\uparrow,\mathrm{out}}(t_{\rm m}))\sin^2(\theta/2)} \\
& \times &
\begin{bmatrix}
    \cos^2(\theta/2)+(1-P_{\uparrow,\mathrm{out}}(t_{\rm m}))\sin^2(\theta/2) & \cos (\theta/2)  \\
    \cos (\theta/2) & 1
\end{bmatrix}. \nonumber
\end{eqnarray}

\normalsize
Hence, the expectation value of $\sigma_z$ as a function of measurement time, conditioned on measuring $\ket{\downarrow}$ (no tunneling) is 
\small
\begin{equation}
\langle \sigma_z (t_{\rm m}) \rangle  = \frac{\cos^2(\theta/2)+\exp(-\Gamma_{\uparrow,\mathrm{out}} t_{\rm m})\sin^2(\theta/2)-1}{\cos^2(\theta/2)+\exp(-\Gamma_{\uparrow,\mathrm{out}} t_{\rm m})\sin^2(\theta/2)+1},
\end{equation}
\normalsize
which for $\theta=\pi$ reduces to a particularly simple form
\begin{equation}
\langle \sigma_z (t_{\rm m}) \rangle  = \frac{\exp(-\Gamma_{\uparrow,\mathrm{out}} t_{\rm m})-1}{\exp(-\Gamma_{\uparrow,\mathrm{out}} t_{\rm m})+1}.
\end{equation}
Solving for $\Gamma_{\uparrow,\mathrm{out}}$ as a function of $\langle \sigma_z (t_{\rm m})\rangle $ we find
\begin{equation} 
\frac{1}{\Gamma_{\uparrow,\mathrm{out}}} = -\frac{t_{\rm m}}{\ln\left(\frac{1+\langle \sigma_z (t_{\rm m})\rangle}{1-\langle \sigma_z (t_{\rm m})\rangle}\right)}. \label{tunnelrate}
\end{equation}

\begin{figure} [tbh]
\includegraphics[width=0.44\textwidth]{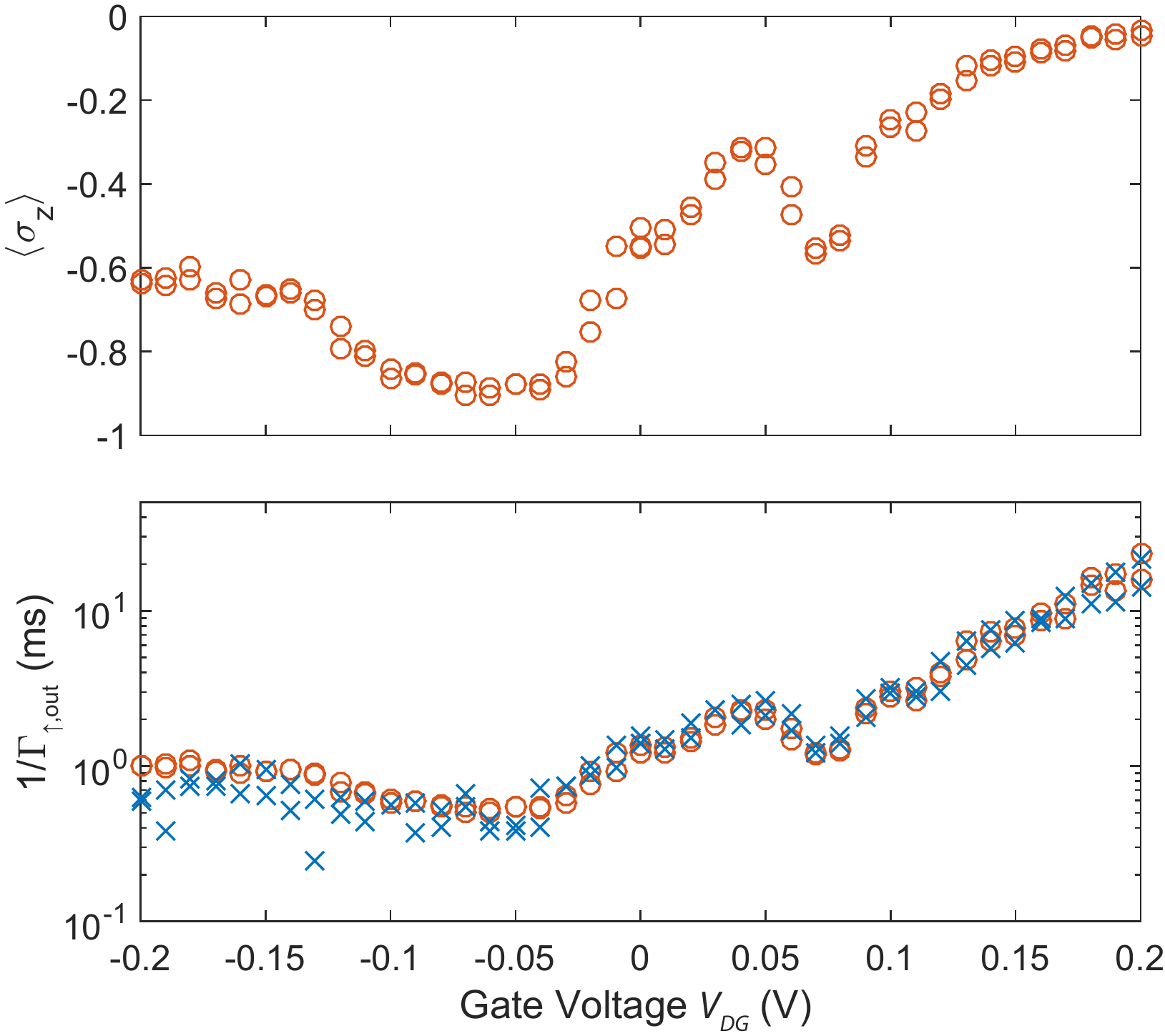}
\caption{Extracting electron tunnel rate from dataset conditioned on having no-tunneling events. (a) Average nuclear polarization $\sigma_z$ after a $t_{\rm m} = 1.5$~ms electron readout window as a function of donor electrochemical potential, controlled by $V_{\rm DG}$. (b) Electron $\ket{\uparrow}$ tunnel-out time $1/\Gamma_{\uparrow,\mathrm{out}}$ extracted from the ``tunnelless'' data in (a) using Eq.~\ref{tunnelrate} (circles), and measured directly from tunneling events (crosses). Data is taken by stepping $V_{\rm DG}$ from low to high value and then in reverse.} 
\label{tunnelmeas}
\end{figure}

In Fig.~\ref{tunnelmeas} we show the results of an experiment where we perform the above mentioned protocol, i.e., we prepare the nucleus in $\ket{\psi_{\rm n0}} = (\ket{\Downarrow} + \ket{\Uparrow})/\sqrt{2}$, the electron in $\ket{\downarrow}$ and then apply an electron $\pi$-pulse at $\nu_{\rm e1}$, thus leaving the electron-nuclear system in the Bell state \cite{Dehollain2016a} $\ket{\Phi^+} = (\ket{\downarrow\Downarrow} + \ket{\uparrow\Uparrow})/\sqrt{2}$. We then bring the electron towards the readout position for a time $t_{\rm m} = 1.5$~ms and, conditional on having no tunneling events, we subsequently measure the nuclear polarization $\langle \sigma_z \rangle$. The experiment is repeated at different values of the gate voltage $V_{\rm DG}$, which controls the donor electrochemical potential $\mu_{\rm D}$ relative to the Fermi level of the electron reservoir \cite{Morello2010}, and thereby tunes the donor-reservoir tunnel rate $\Gamma_{\uparrow,\mathrm{out}}$. For $V_{\rm DG} \gtrsim 0.2$~V the $\ket{\uparrow}$ state is well below the Fermi level and neither the $\ket{\downarrow}$ nor the $\ket{\uparrow}$ states have a significant probability of tunneling out, i.e. the measurement strength vanishes: the absence of a tunneling event does not imply a $\ket{\downarrow}$ state. Accordingly, we find $\langle \sigma_z \rangle \approx 0$ in that limit, i.e. the nuclear polarization has not been perturbed from the initial value. For $V_{\rm DG} < 0.2$~V, $1/\Gamma_{\uparrow,\mathrm{out}}$ becomes shorter and $\langle \sigma_z \rangle$ veers towards negative values, which indicates that the electron $\ket{\downarrow}$ measurement is becoming stronger, thus turning the initial $\ket{\Phi^+}$ Bell state towards $\ket{\downarrow\Downarrow}$. Using Eq.~\ref{tunnelrate} we can extract the numerical value of $1/\Gamma_{\uparrow,\mathrm{out}}$, and compare it [Fig.~\ref{tunnelmeas}(b)] to the tunnel time extracted directly from tunneling probabilities. The two methods agree almost perfectly, confirming the validity of our approach. The non-monotonic behavior of $\Gamma_{\uparrow,\mathrm{out}}(V_{\rm DG})$ is related to modulations in the density of states of the electron reservoir \cite{Mottonen2010}.

Unlike the weak nuclear measurement describe earlier, this process using weak electron measurement does not preserve the purity of the nuclear spin state. Also, the use of a maximally entangled $\ket{\Phi^+}$ Bell state as the starting point of the sequence is inconsequential for this particular experiment -- the same result would be obtained starting from an incoherent mixture of \ket{\downarrow\Downarrow} and \ket{\uparrow\Uparrow}, though the perfect correlation between the two spins is obviously required. Nonetheless, the process provides a curious example of interaction-free measurement \cite{Elitzur1993} in the solid state.

In conclusion, we have shown the application of several concepts and tools of weak single-shot measurements to a model solid-state spin system. In particular, we have demonstrated the ability to \emph{coherently} control a nuclear spin using only ESR pulses and electron spin readout, and we have shown how to measure tunnel rates without any tunneling events. In the future, these techniques can be applied to a variety of interesting problems, such as the study of qubit dynamics under driving and weak measurement \cite{Ruskov2007}, past quantum states of a monitored system \cite{Gammelmark2013} and the use of steering to improve qubit initialization. 

\acknowledgments{We thank K. M{\o}lmer and R. Ruskov for insightful comments. This research was funded by the Australian Research Council through a Discovery Project (DP150101863) and the Centre of Excellence Quantum Computation and Communication Technology (CE11E0001027), the US Army Research Office (W911NF-13-1-0024) and the Commonwealth Bank of Australia. We acknowledge support from the Australian National Fabrication Facility, and from the laboratory of Prof Robert Elliman at the Australian National University for the ion implantation facilities. The work at Keio has been supported in part by KAKENHI (S) No. 26220602, Core-to-Core Program by JSPS, and Spintronics Research Network of Japan.}  


%

\newpage

\widetext{
\ \\
\ \\
\textbf{SUPPLEMENTARY MATERIAL: Coherent control via weak measurements in $^{31}$P single-atom electron and nuclear spin qubits}

\section{Success probabilities}

Performing a conditional weak measurement is necessarily a probabilistic process. As mentioned in the main text, the success probability for a single measurement [starting from the nuclear spin state in equation (1) of the main text] is $P_1 = [1+\cos(\theta/2)]/2$. It is however notable that, as this probability depends on the nuclear spin populations at the start of the measurement, the success probability of two sequential weak measurements is not simply this value squared. Rather, the success probability for $n$ sequential weak measurements in our case is $P_n = [1+\cos(\theta/2)^{2n}]/2$ if all measurements are performed with electron spin rotation $\theta$ on the same electron spin resonance frequency.

For the measurement reversal (two weak nuclear measurements, each using a different ESR frequency) the success probability reads $P_{rev} = \cos^2(\theta/2)$ which is notably zero for $\theta=\pi$, as should be expected (one cannot reverse a projective measurement). These predictions together with data are plotted in Fig.~4.

\begin{figure}[h!]
\includegraphics[width=0.6\textwidth]{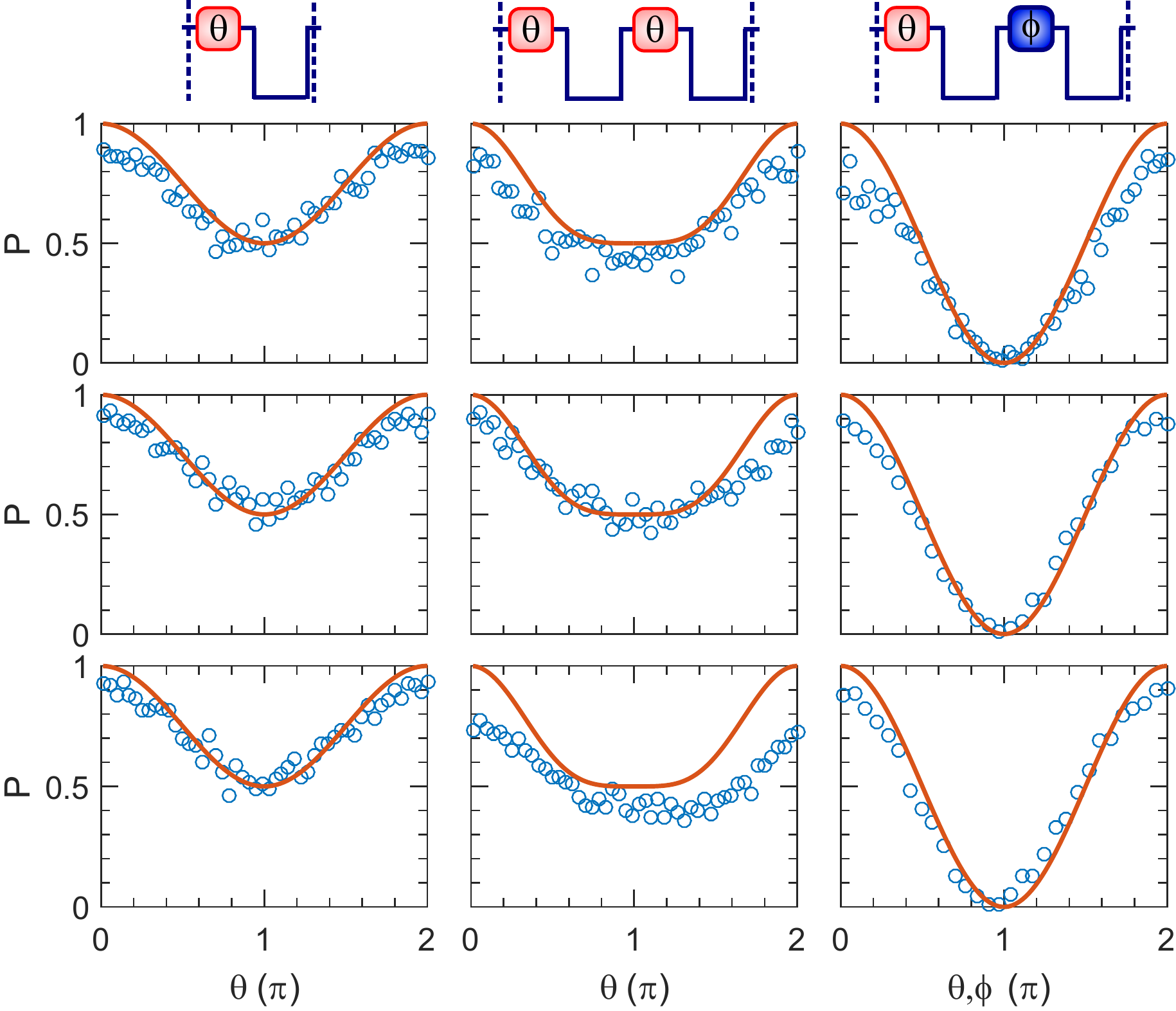}
\caption{Probabilities of successful conditional measurement. Same dataset as main figure 2. First row shows the measurements for $\sigma_z$ component, second row the $\sigma_x$ component data and third row the $\sigma_y$ component data. Success probability should not depend on the tomography axis but we show all data for completeness. Columns are from left to right: one measurement, two measurements (each with rotation angle $\theta$) and measurement reversal (rotation $\theta$,$\phi$ on different ESR frequencies). Circles show experimental data and lines are theoretical predictions with no fitting parameters. Each point corresponds to 200 repetitions.}
\end{figure}

\section{Density matrix calculations}

Below we refer to the nuclear spin state with the thick arrow ($\Uparrow$ or $\Downarrow$) and the electron spin state with the narrow arrow ($\uparrow$ or $\downarrow$). The Pauli operators are $\sigma^i$ where $i=e,n$ refers to either electron or nuclear spin, respectively. We start from the state $\Phi = 1/\sqrt{2}(\ket{\Uparrow}+\ket{\Downarrow})\otimes\ket{\downarrow}$, i.e., in density matrix form (in the basis  $\ket{\Uparrow\uparrow}\ket{\Uparrow\downarrow}\ket{\Downarrow\uparrow}\ket{\Downarrow\downarrow}$)
\begin{equation}
\rho_0 = \frac{1}{2}
\begin{bmatrix}
    0 & 0 & 0 & 0  \\
    0 & 1 & 0 & 1  \\
    0 & 0 & 0 & 0 \\
    0 & 1 & 0 & 1\\
\end{bmatrix}.
\end{equation}
The conditional rotation matrix reads $U(\theta) = \ket{\Downarrow}\bra{\Downarrow}\otimes I + \ket{\Uparrow}\bra{\Uparrow} \otimes R(\theta)$ where $R(\theta)$ is the rotation matrix
\begin{equation}
R(\theta) =
\begin{bmatrix}
    \cos(\theta) & -\sin(\theta) \\
    \sin(\theta) & \cos(\theta) \\
\end{bmatrix}.
\end{equation}
Hence, after the initialization step and the conditional electron spin rotation of an angle $\theta$ the system state is
\begin{equation}
\rho_\theta = U(\theta)\rho_0 U^\dag(\theta) = \frac{1}{2}
\begin{bmatrix}
    \sin^2(\theta/2) & \cos(\theta/2)\sin(\theta/2) & 0 & \sin(\theta/2)  \\
    \cos(\theta/2)\sin(\theta/2) & \cos^2(\theta/2) & 0 & \cos(\theta/2)  \\
    0 & 0 & 0 & 0 \\
    \sin(\theta/2) & \cos(\theta/2) & 0 & 1\\
\end{bmatrix}.
\end{equation}
which is an entangled electron-nuclear state for all $\theta \neq 0,2\pi$ (according to the PPT criterion).

If we then just simply trace out the electron (no conditioning), we obtain the nuclear spin state as
\begin{equation}
\rho_n^u = \Tr_2(\rho_\theta) =  \frac{1}{2}
\begin{bmatrix}
    \sin^2(\theta/2) + \cos^2(\theta/2) & \cos(\theta/2) \\
    \cos(\theta/2) & 1  \\
\end{bmatrix}
= \frac{1}{2}
\begin{bmatrix}
    1 & \cos(\theta/2) \\
    \cos(\theta/2) & 1  \\
\end{bmatrix},
\end{equation}
showing that the expectation value of $\sigma_z^n$ remains constant independently of $\theta$, but the off-diagonal elements decay as a function of the measurement strength. In the limiting case of $\theta=\pi$, we are left with a classical mixture of up and down nuclear spin states.

More interestingly, tracing out the electron conditionally on measuring \ket{\downarrow} we obtain
\begin{equation}
\rho_n^c = \textrm{Tr}_2\left[\rho_\theta \left(\textrm{I} \otimes \ket{0}\bra{0}\right)\right]
= \frac{1}{1+\cos^2(\theta/2)}
\begin{bmatrix}
    \cos^2(\theta/2) & \cos(\theta/2) \\
    \cos(\theta/2) & 1,  \\
\end{bmatrix}
\end{equation}
which is the state mentioned in the main text. The second measurement is then simply done by repeating the process starting from this state
\begin{equation}
\rho_\theta^{(2)} = U(\theta)\left(\rho_n^c\otimes\ket{\downarrow}\right) U^\dag(\theta)
\end{equation}
and tracing out similarly. For the measurement reversal, we need the rotation matrix for the other electron spin resonance frequency, which reads  $U(\theta) = \ket{\Downarrow}\bra{\Downarrow}\otimes R(\theta) + \ket{\Uparrow}\bra{\Uparrow} \otimes I$. Otherwise the procedure is the same. Expectation values for the nuclear spin components for one or two measurements with rotation $\theta$ are plotted in Fig.~5. The measurement reversal should just preserve all three components. These are also plotted with the data in main figure 2.

\begin{figure}
\includegraphics[width=0.9\textwidth]{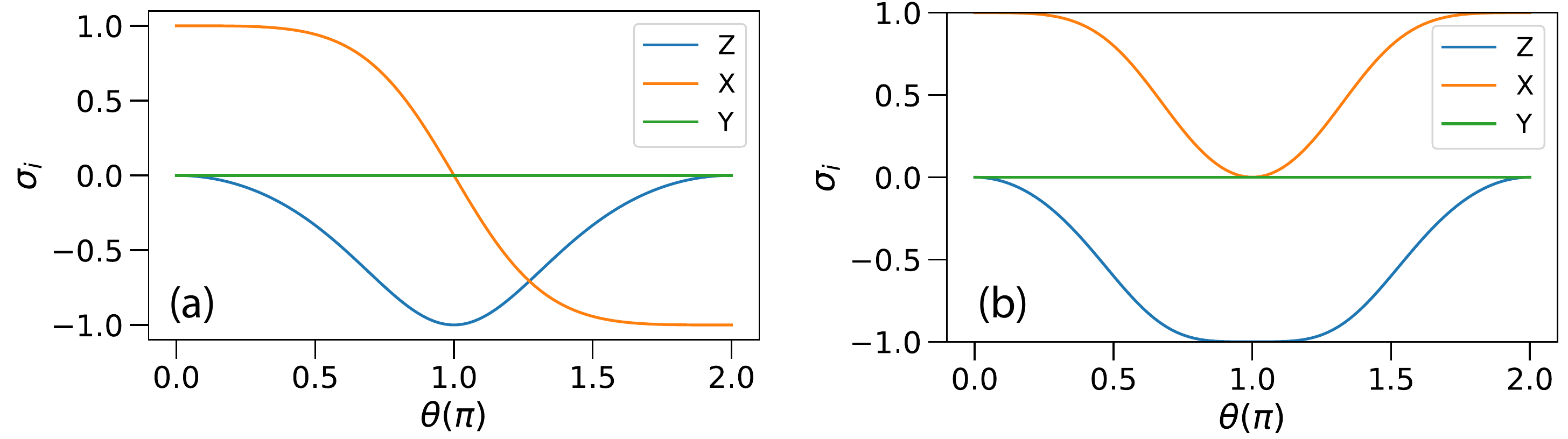}
\caption{
Expectation values for the nuclear spin components after one (a) or two (b) conditional weak measurements as a function of the rotation angle $\theta$ (measurement strength). In (b) both measurements are performed with the same $\theta$. 
}
\end{figure}

Finally, if we also add a finite electron tunnel-out probability to process described above, we obtain
\begin{eqnarray}
\rho_n^c &=& \textrm{Tr}_2 \left\lbrace  \rho_\theta \left[ \textrm{I} \otimes \left( \ket{0}\bra{0}+\exp(-\Gamma t)\ket{1}\bra{1} \right) \right] \right\rbrace  \nonumber \\ 
&=& \frac{1}{1+\cos^2(\theta/2)+\exp(-\Gamma t)\sin^2(\theta/2)}
\begin{bmatrix}
    \cos^2(\theta/2)+\exp(-\Gamma t)\sin^2(\theta/2) & \cos(\theta/2) \\
    \cos(\theta/2) & 1  \\
\end{bmatrix}.
\end{eqnarray}
Note that, unlike all the previous states, this one is not pure unless $\exp(-\Gamma t)\sin^2(\theta/2) = 0$.

\section{Notes on EPR steering}

The use of the word ``steering'' in the context of quantum systems is somewhat ambigous in the existing literature. The experiments in this paper demonstrate coherent control of a qubit state by measuring another, correlated, qubit state. This is in many contexts called steering and this usage of the word indeed makes intuitive sense; one is steering the nuclear spin (qubit) by weakly measuring it via the electron (ancilla).

However, it is also common that the word steering - in the quantum context - exclusively refers to what is more exactly known as EPR steering. In the operational definition of Wiseman \emph{et al}., EPR steering consists of a ``game'' where Alice must convince Bob that she has shared with him an entangled state. To do so, she wants to show Bob that she has the ability to control his quantum state by choosing which measurement to perform at her end. This, in turn, can be formalized in experimentally testable EPR steering inequalities.  

A demonstration of EPR steering could be conducted on the $^{31}$P electron-nuclear system, where ``Alice'' is the electron spin and ``Bob'' is the nuclear spin, by following three steps:

(i) Initialize the electron-nuclear system in a maximally entangled Bell state, for example $\ket{\Phi^+} = (\ket{\downarrow\Downarrow} + \ket{\uparrow\Uparrow})/\sqrt{2}$, as described in the main text. 

(ii) Define different measurement axes for the electron spin. This requires an \emph{unconditional} electron spin rotation, which could be obtained by simultaneously applying ESR pulses of rotation angle $\theta$ on both $\nu_{\rm e1}$ and $\nu_{\rm e2}$, before a projective electron spin measurement. This is the key difference between EPR steering and the experiments shown in the main text, where all electron spin rotations were \emph{conditional} on the nuclear spin state. This is because simultaneous excitation of $\nu_{\rm e1}$ and $\nu_{\rm e2}$ was not feasible in our setup. 

(iii) Conditioned on measuring electron spin \ket{\downarrow}, perform nuclear state tomography. 

Supplementary figure~6 shows the expected nuclear spin components as a function of $\theta$. At $\theta = 0$ the electron spin measurement is along the $z$-axis and therefore the subsequent measurement of $\sigma_z^n$ could be predicted by unity accuracy, whereas the measurement of $\sigma_x^n$ is completely undetermined. At $\theta = \pi/2$ the electron spin measurement is along the $x$-axis, and now the reverse is true. This simple simulation captures the essence of EPR steering. The state of Bob's particle tracks exactly the choice of measurement basis made by Alice.

We note that the violation of Bell's inequality has already been demonstrated with the electron-nuclear system studied here, and it is known that the requirements for EPR-steering are less strict than those for Bell inequalities. Therefore, using e.g. two separate microwave sources to excite $\nu_{\rm e1}$ and $\nu_{\rm e2}$ simultaneously, it should be possible to demonstrate EPR steering in the $^{31}$P system.

\begin{figure}
\includegraphics[width=0.45\textwidth]{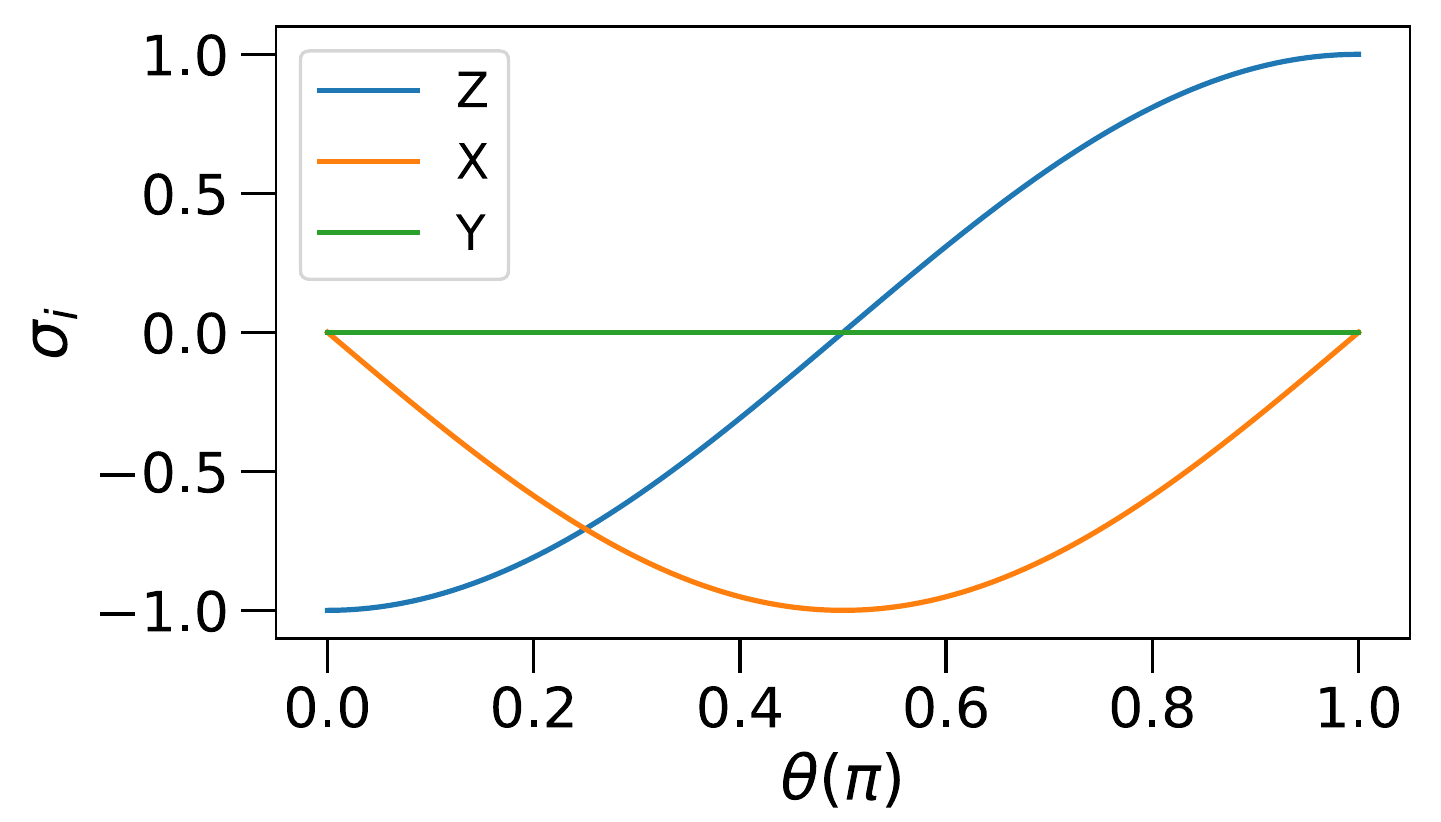}
\caption{EPR-steering. Expectation values for the nuclear spin component, after one starts from the Bell state, does an unconditional rotation of the electron spin with angle $\theta$ and then a conditional measurement of the electron. As the unconditional rotation changes the electron measurement basis, we have a perfect correlation between measuring electron down and nuclear spin Z-component at $\theta=0,\pi$. At $\theta=\pi/2$ we have unity correlation with the $\sigma_x^n$ component.
}
\end{figure}

\end{document}